\documentclass[12pt]{article}
\usepackage[utf8]{inputenc}
\usepackage{authblk}
\usepackage{setspace}
\usepackage[margin=1.25in]{geometry}
\usepackage{graphicx}
\usepackage{subcaption}
\usepackage{amsmath}
\usepackage{lineno}
\usepackage{lipsum}
\usepackage[style=nejm, 
citestyle=numeric-comp,
sorting=none]{biblatex}
\begin{document}
\title{Measurement of leading charged-particle jet properties in p--Pb collisions at $\sqrt{s_{\rm NN}}$~=~5.02~TeV with ALICE}
\author[1*]{Prottoy Das (for the ALICE Collaboration)}
\affil[1]{Bose Institute, Kolkata, India}
\affil[*]{Address correspondence to: prottoy.das@cern.ch}

\onehalfspacing
\maketitle
\date{}
\begin{abstract}
        Jets are collimated sprays of particles produced from the fragmentation and hadronization of hard-scattered partons in high energy hadronic and nuclear collisions. Jet properties are sensitive to details of parton showering processes and are expected to be modified in the presence of a dense partonic medium. Measurement of intra-jet properties in p--Pb collisions will help to investigate cold nuclear matter effects and enrich our current understanding of particle production in such collision systems. In this work, we present the measurement of leading charged-particle jet properties, namely the mean charged-particle multiplicity and the fragmentation functions, in the range of jet $p_{\rm T}$ 10~--~100 GeV/c at midrapidity in minimum bias p--Pb collisions at $\sqrt{s_{\rm NN}}$~=~5.02~TeV with ALICE. Results are compared with theoretical model predictions.
\end{abstract}
\newpage
Measurements of intra-jet properties in p--Pb collisions are useful to investigate cold nuclear matter effects~\cite{CNM}, constrain theoretical models, and enrich our current understanding of particle production in such collision systems. The analysis was carried out on a sample of p--Pb collisions at $\sqrt{s_{\rm NN}}$~=~5.02~TeV collected with ALICE~\cite{ALICE_det} at the LHC using a minimum bias trigger condition that requires the coincidence of signals in the V0A and V0C forward scintillator arrays. The accepted events are also required to have a primary vertex within $\pm$10 cm from the nominal interaction point along the beam direction. Charged tracks are reconstructed using information from the Inner Tracking System (ITS) and the Time Projection Chamber (TPC)  with minimum $p_{\rm T}$~=~0.15 GeV/$c$ in the pseudorapidity range $|\eta|<$~0.9 over the full azimuth. Charged-particle jets are reconstructed from the selected tracks using the anti-$k_{\rm T}$ jet finding algorithm of FastJet 3.2.1~\cite{FastJet} for jet resolution parameter $R$~=~0.4. The mean charged-particle multiplicity in jet ($\langle N_{\rm ch} \rangle$) and jet fragmentation function (${\rm d}N/{\rm d}z^{\rm ch}$ with $z^{\rm ch}=p_{\rm T, track}/p_{\rm T,jet}^{\rm ch}$, where $p_{\rm T, track}$ is the $p_{\rm T}$~of jet constituent) for leading jets (jet with the highest $p_{\rm T}$ in an event) with 10~$<p_{\rm T, jet}^{\rm ch}<$~100~GeV/$c$ are studied as a function of jet $p_{\rm T}$. A two-dimensional Bayesian unfolding technique~\cite{Bayesian} (implemented in the RooUnfold~\cite{RooUnfold} package) is applied to correct for the instrumental effects. The underlying event (UE) contribution is estimated using the perpendicular cone method and subtracted on a statistical basis after unfolding. The dominant systematic uncertainty is from tracking inefficiency. More details of the analysis can be found in Ref.~\cite{JetProppPb}.

\begin{figure}
\vspace{-1.3cm}   
        \begin{subfigure}[b]{0.34\linewidth}
		\includegraphics[width=\columnwidth]{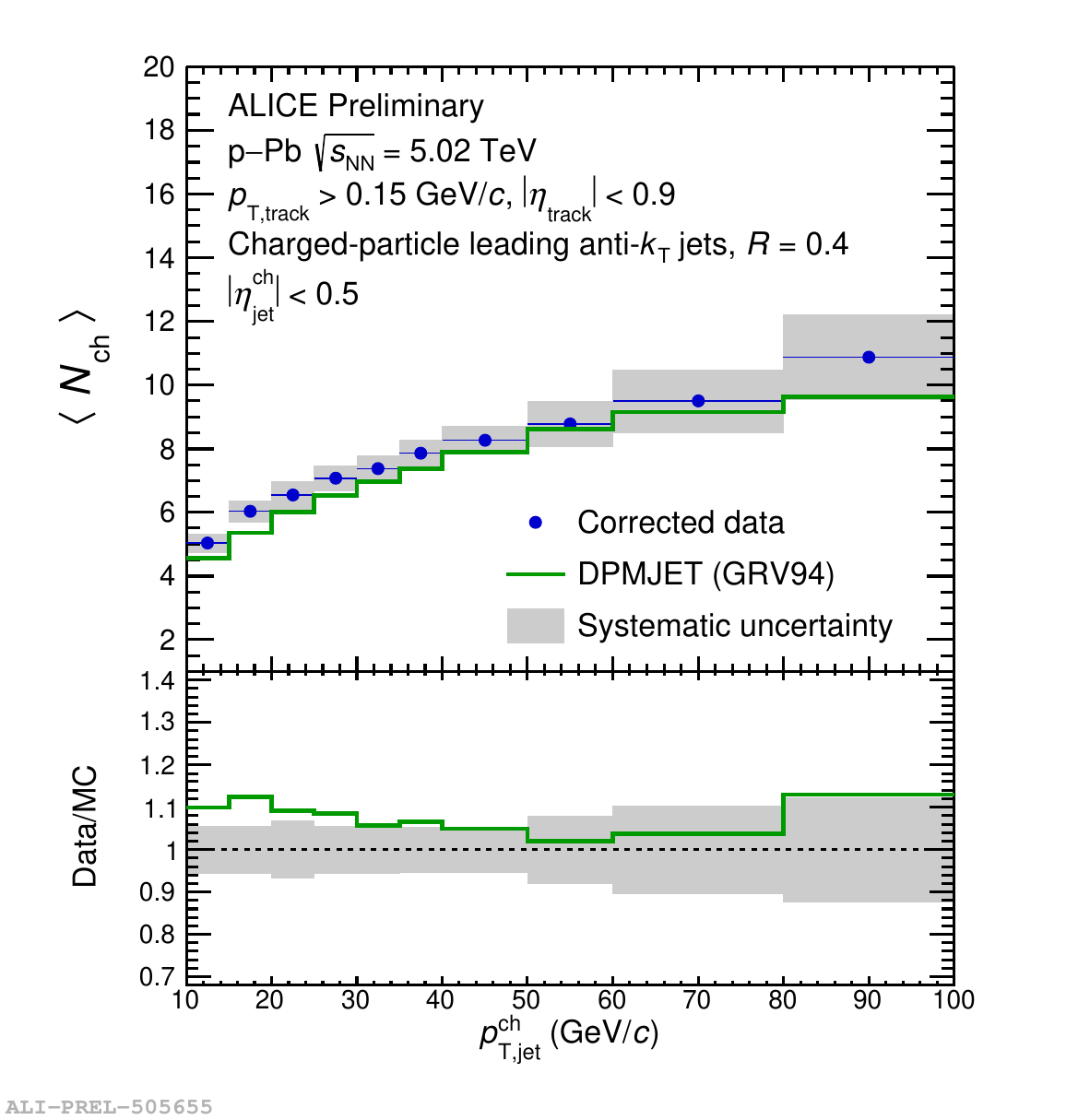}                
		\caption{}
		\label{Fig:NchMB}
	\end{subfigure}
        \hspace{-0.6cm}
	\begin{subfigure}[b]{0.34\linewidth}
		\includegraphics[width=\columnwidth]{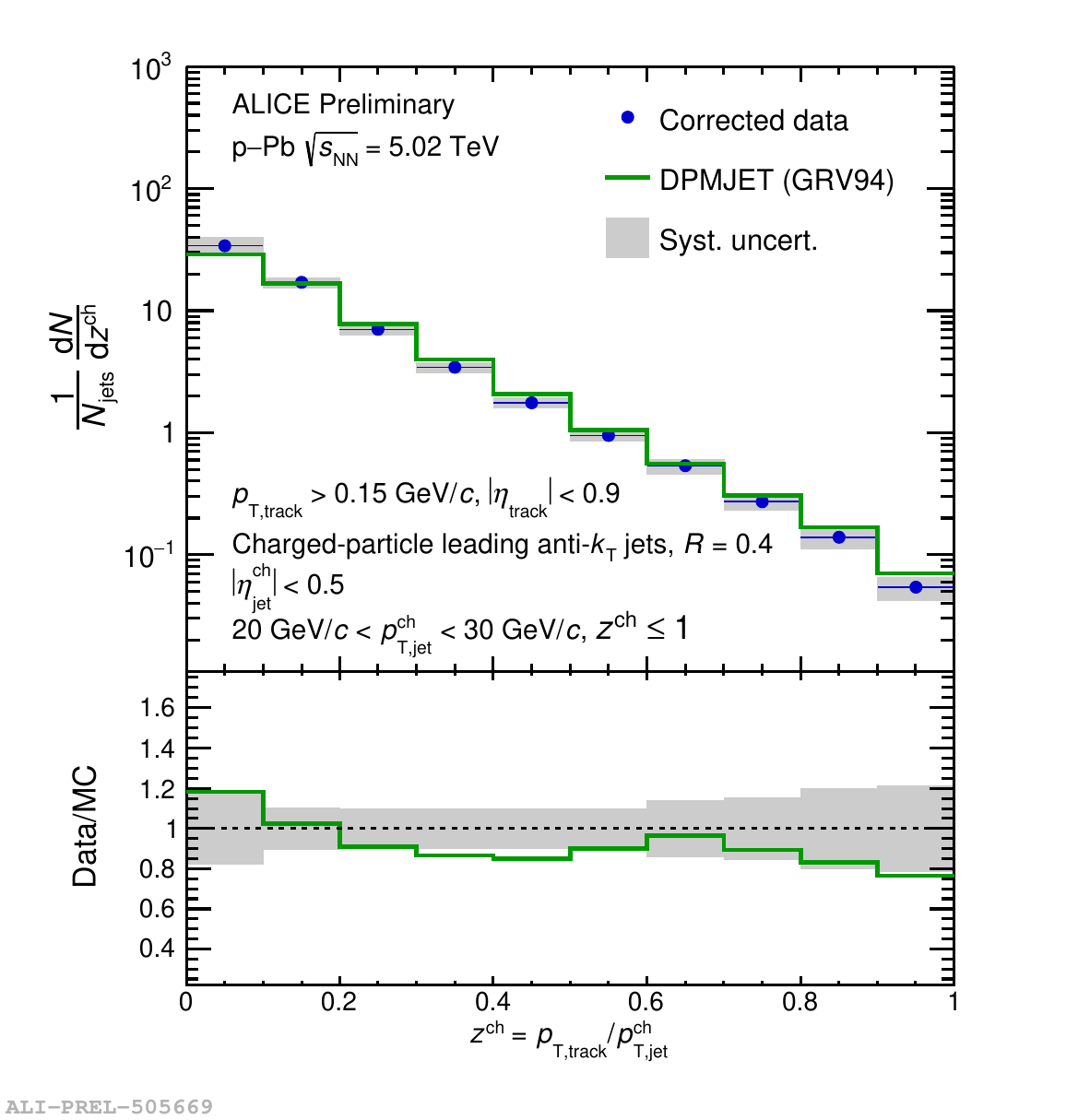}
		\caption{}
		\label{Fig:FFMB20to30}
	\end{subfigure}
        \hspace{-0.6cm}
        \begin{subfigure}[b]{0.37\linewidth}
                \vspace{-0.13cm}         
		\includegraphics[width=\columnwidth]{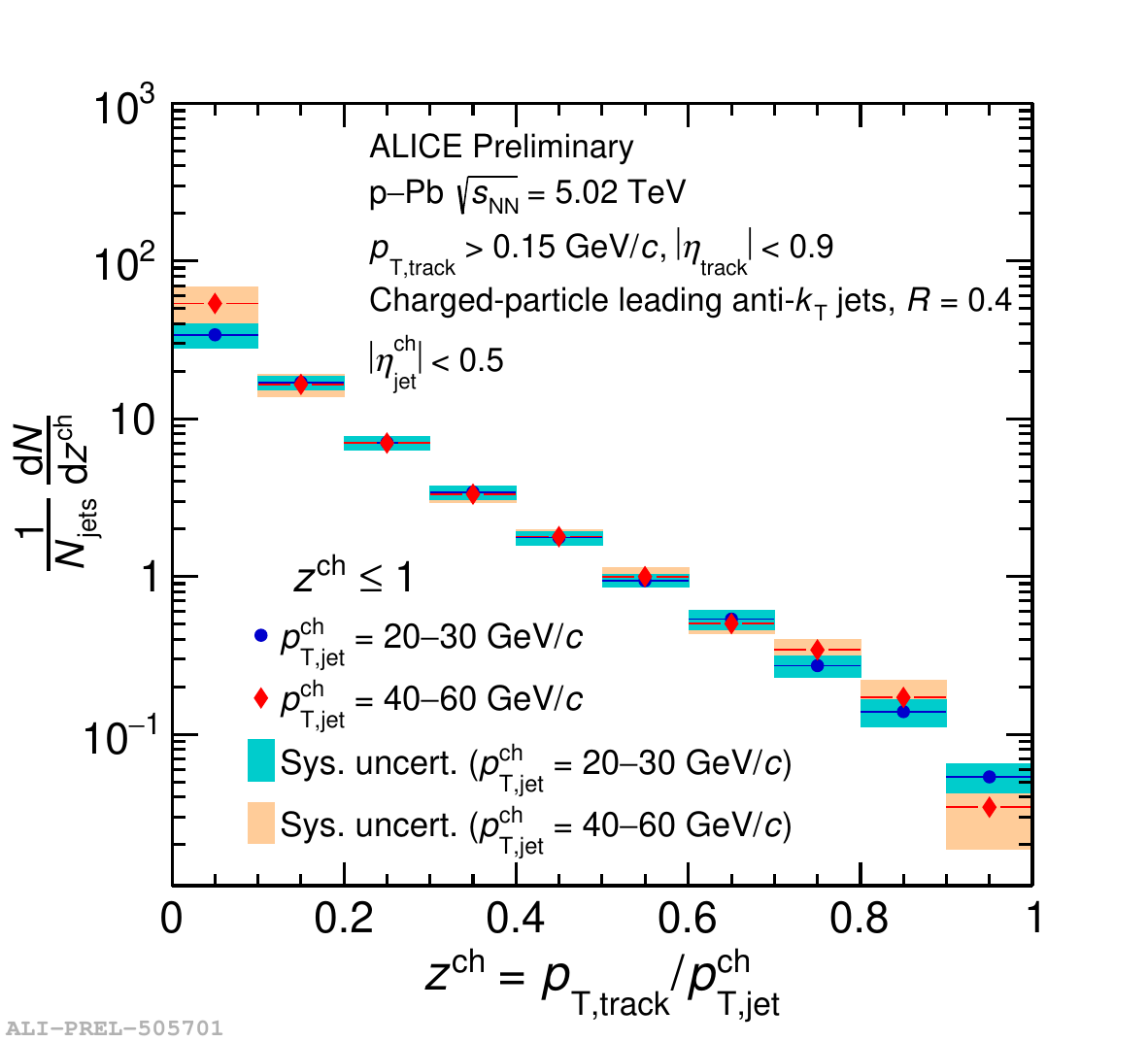}
		\caption{}
		\label{Fig:FFAllMB}
	\end{subfigure}
        \vspace{-0.25cm}
        \caption{Top panels: (a) $\langle N_{\rm ch}\rangle$ distribution as a function of leading jet $p_{\rm T, jet}^{\rm ch}$. (b) $z^{\rm ch}$ distributions for leading jet $p_{\rm T, jet}^{\rm ch}$~=~20~--~30 GeV/${c}$. Bottom panels: ratios between the data and DPMJET predictions.\\ (c) $z^{\rm ch}$ distributions for leading jet $p_{\rm T, jet}^{\rm ch}$~=~20~--~30 GeV/${c}$ and 40~--~60 GeV/${c}$.}
         \vspace{-0.4cm}
 \end{figure}

Figures~\ref{Fig:NchMB} and~\ref{Fig:FFMB20to30} show the $\langle N_{\rm ch}\rangle$ distribution as a function of leading jet $p_{\rm T, jet}^{\rm ch}$ and the $z^{\rm ch}$ distribution for leading jets with 20~$<p_{\rm T, jet}^{\rm ch}<$~30~GeV/$c$, respectively. The distributions are compared to DPMJET~\cite{DPMJET} model predictions and the bottom panels of the figures display the data-to-DPMJET ratios. It can be seen that $\langle N_{\rm ch}\rangle$ increases with $p_{\rm T, jet}^{\rm ch}$ and DPMJET (GRV94) describes the $\langle N_{\rm ch}\rangle$ distribution for $p_{\rm T, jet}^{\rm ch}>$ 30 GeV/$c$ and the $z^{\rm ch}$ distribution within systematic uncertainties. The comparison of the $z^{\rm ch}$ distributions between two different $p_{\rm T, jet}^{\rm ch}$ ranges, as shown in Fig.~\ref{Fig:FFAllMB}, indicates that the $z^{\rm ch}$ distribution follows a scaling behaviour independent of $p_{\rm T, jet}^{\rm ch}$ within systematic uncertainties. This implies that the probability of jet constituents having a given fraction of $p_{\rm T, jet}^{\rm ch}$ is independent of the total jet $p_{\rm T, jet}^{\rm ch}$ in this kinematic range.

\printbibliography

\end{document}